\documentclass[12pt, draftclsnofoot, onecolumn]{IEEEtran}
\usepackage{amsmath,amssymb}
\usepackage{amsthm}

\usepackage[inline]{enumitem}

\usepackage{graphicx}
\usepackage[font=small]{caption}

\usepackage{cite}

\ifCLASSOPTIONcompsoc
\usepackage[caption=false,font=normalsize,labelfont=sf,textfont=sf]{subfig}
\else
\usepackage[caption=false,font=footnotesize]{subfig}
\fi

\usepackage{algorithm}
\usepackage{algorithmic}

\usepackage{color}

\makeatletter
\def\blfootnote{\gdef\@thefnmark{}\@footnotetext}
\makeatother

\newcommand{\mat}[1]{\begin{bmatrix} #1 \end{bmatrix}}

\newcommand{\norm}[1]{\left\| #1 \right\|}
\newcommand{\abs}[1]{\left| #1 \right|}
\newcommand{\Ev}{\mathbb{E}}
\newcommand{\trace}{\mathrm{Tr}}
\newcommand{\herm}{\mathrm{H}}
\newcommand{\tr}{\mathrm{T}}
\newcommand{\inv}{{-1}}
\newcommand{\SNR}{\mathrm{SNR}}

\newcommand{\Rsum}{R_\mathrm{sum}}

\newcommand{\IntF}{\mathrm{IF}}
\newcommand{\DIF}{\mathrm{DIF}}

\newcommand{\HI}{\mathrm{HI}}
\newcommand{\Rcomp}{R_{\textrm{\textup{comp}}}}

\DeclareMathOperator{\diag}{diag}

\DeclareMathOperator*{\minimize}{minimize}
\DeclareMathOperator{\rank}{rank}

\newtheorem{theorem}{Theorem}

\newcommand{\tilA}{\tilde{\bfA}}
\newcommand{\tilD}{\tilde{\bfD}}

\newcommand{\ZZ}{\mathbb{Z}}

\newcommand{\RR}{\mathbb{R}}
\newcommand{\CC}{\mathbb{C}}

\newcommand{\bLambda}{\mathbf{\Lambda}}
\newcommand{\bNabla}{{\pmb \nabla}}

\newcommand{\calA}{\mathcal{A}}

\newcommand{\calC}{\mathcal{C}}

\newcommand{\calN}{\mathcal{N}}
\newcommand{\calO}{\mathcal{O}}

\newcommand{\calW}{\mathcal{W}}

\newcommand{\bfA}{\mathbf{A}}

\newcommand{\bfD}{\mathbf{D}}

\newcommand{\bfH}{\mathbf{H}}
\newcommand{\bfI}{\mathbf{I}}

\newcommand{\bfM}{\mathbf{M}}

\newcommand{\bfP}{\mathbf{P}}

\newcommand{\bfT}{\mathbf{T}}
\newcommand{\bfU}{\mathbf{U}}

\newcommand{\bfa}{\mathbf{a}}

\newcommand{\bfh}{\mathbf{h}}

\newcommand{\bfw}{\mathbf{w}}

\newcommand{\bfy}{\mathbf{y}}

\newcommand{\bA}{\mathbf{A}}
\newcommand{\bB}{\mathbf{B}}

\newcommand{\bD}{\mathbf{D}}

\newcommand{\bH}{\mathbf{H}}
\newcommand{\bI}{\mathbf{I}}

\newcommand{\bL}{\mathbf{L}}
\newcommand{\bM}{\mathbf{M}}

\newcommand{\bP}{\mathbf{P}}

\newcommand{\bT}{\mathbf{T}}
\newcommand{\bU}{\mathbf{U}}

\newcommand{\bX}{\mathbf{X}}
\newcommand{\bY}{\mathbf{Y}}
\newcommand{\bZ}{\mathbf{Z}}

\newcommand{\ba}{\mathbf{a}}

\newcommand{\bh}{\mathbf{h}}

\newcommand{\bp}{\mathbf{p}}

\newcommand{\bw}{\mathbf{w}}
\newcommand{\bx}{\mathbf{x}}
\newcommand{\by}{\mathbf{y}}
\newcommand{\bz}{\mathbf{z}}

\newcommand{\bfzero}{\mathbf{0}}

\begin{document}

\title{Optimization of Integer-Forcing Precoding for Multi-User MIMO Downlink}

\author{Ricardo Bohaczuk Venturelli and Danilo Silva,~\IEEEmembership{Member,~IEEE}%
	\thanks{Manuscript received ?????? ??, ????; revised ?????? ??, ????; accepted
		?????? ??, ????. Date of publication ?????? ??, ????; date of current
		version ?????? ??, ????. 
		This work was supported by CNPq, Brazil under Grant 153535/2016-4, Grant 310343/2016-0, and Grant 429097/2016-6.
		The associate editor coordinating the review of this
		paper and approving it for publication was ???? ????. (Corresponding author:
		Ricardo Bohaczuk Venturelli.)}
	\thanks{Ricardo Bohaczuk Venturelli and Danilo Silva are with the Department of Electrical and Eletronic Engineering, Federal University of Santa Catarina, Florian\'{o}polis-SC 88040-900, Brazil (e-mails: ricardo.bventurelli@gmail.com, danilo.silva@ufsc.br).}
	\thanks{Digital Object Identifier ??????}%
\thanks{This work has been submitted to the IEEE for possible publication.  Copyright may be transferred without notice, after which this version may no longer be accessible.}}
\maketitle

\begin{abstract}
Integer-forcing (IF) precoding is an alternative to linear precoding for multi-user (MU) multiple-input-multiple-output (MIMO) channels, with the potential to offer superior performance at a similar complexity. In this letter, a low-complexity suboptimal method is proposed to optimize the parameters of an IF scheme for any number of $K$ users. The proposed method involves solving a relaxation of the problem followed by the application of a lattice reduction algorithm and is shown to have an overall complexity of $\calO(K^3)$. Simulation results show that the proposed method achieves a higher sum rate than a heuristic choice of parameters and significantly outperforms conventional linear precoding in all simulated scenarios.
\end{abstract}

\section{Introduction} \label{sec:introduction}

Precoding techniques are often used in order to mitigate user interference in multi-user (MU) multiple-input-multiple-output (MIMO) downlink channels \cite{Tse.2005:FundamentalsWirelessCommunication}. 
Linear precoding methods, such as zero-forcing (ZF) and regularized ZF (RFZ) \cite{Bjornson.2014:Optimalmultiusertransmit,Peel.2005:vectorperturbationtechniquenearcapacitya}, are widely used due to their low complexity, however, their performance falls far below the sum capacity. On the other hand, non-linear techniques, such as vector-pertubation \cite{Peel.2005:vectorperturbationtechniquenearcapacitya,Hochwald.2005:vectorperturbationtechniquenearcapacitya,Avner.2015:vectorperturbationprecoding}, can achieve higher sum rates in exchange for potentially much higher computational cost.

Lattice-reduction-aided (LRA) precoding \cite{Windpassinger.2004:Latticereductionaidedbroadcastprecodinga,Stern.2016:Advancedfactorizationstrategies} is a low-complexity non-linear technique that, differently from linear methods, can achieve full diversity supported by the channel. 
In LRA precoding, a linear precoding matrix~$\bT$ is applied before transmission, in order to transform the channel matrix~$\bH$ to a more suitable basis (according to some heuristic), which is obtained through lattice basis reduction \cite{Stern.2016:Advancedfactorizationstrategies}. With this approach, the effective channel matrix, after appropriate scaling by the users, becomes a (unimodular) integer-valued matrix $\bA$. In order to cancel this integer interference, prior to the application of~$\bT$, the modulation symbols are pre-multiplied by the inverse of~$\bA$, followed by a modulo operator to limit the transmit power. Since channel coding can be applied on top of an LRA precoding scheme, the performance of the latter is typically measured based on uncoded symbol error probability \cite{Windpassinger.2004:Latticereductionaidedbroadcastprecodinga,Stern.2016:Advancedfactorizationstrategies}.

A generalization of LRA precoding is the so-called integer-forcing (IF) precoding \cite{Hong.2012:Reversecomputeforward,He.2018:UplinkDownlinkDualityIntegerForcing,Stern.2016:Advancedfactorizationstrategies,Silva.2017:integerforcingprecodingGaussian}, whose main difference is that channel encoding is applied immediately before the multiplication by~$\bT$. This approach has the advantage of providing higher reliability at a similar computational cost. Moreover, it allows achievable rate expressions to be derived explicitly, rather than evaluated by numerical simulation as in LRA precoding, leading to a scheme much more amenable to optimization.

However, \textit{optimal} IF precoding (as well as optimal linear precoding) is NP-hard in general \cite{Silva.2017:integerforcingprecodingGaussian} and for this reason prior work has focused on developing low-complexity suboptimal algorithms. The simplest approach is to choose~$\bT$ such that $\bH\bT = c\bA$ \cite{Hong.2012:Reversecomputeforward} or $\bH\bT \approx c\bA$ \cite{Stern.2016:Advancedfactorizationstrategies}, 	where $c>0$ is some constant. This turns out to be equivalent to the LRA approach to choosing $\bT$, requiring lattice reduction to find $\bA$ \cite{Stern.2016:Advancedfactorizationstrategies}. A more general but much more complex approach is the iterative duality-based algorithm in \cite{He.2018:UplinkDownlinkDualityIntegerForcing}, which requires a lattice reduction step at every iteration.\footnote{Another difficulty with the approach of \cite{He.2018:UplinkDownlinkDualityIntegerForcing} is that it requires a more complicated transmission scheme using multiple shaping lattices, so in effect it cannot be applied to the problem considered in this paper.} In \cite{Silva.2017:integerforcingprecodingGaussian}, Silva \textit{et al.} show that, for high SNR, the optimal choice of $\bT$ satisfies $\bH\bT = c\bD\bA$, where $\bD$ is a diagonal matrix, while, for general SNR, the performance can be improved by choosing $\bH\bT \approx c\bD\bA$. For the special case of $K=2$ users, at high SNR, the optimal choice of~$\bD$ and~$\bA$ is found analytically in \cite{Silva.2017:integerforcingprecodingGaussian}, however, the general case remains open.
	
In this letter, we propose a low-complexity sub-optimal method for choosing $\bD$ and $\bA$ for any number of $K$ users. We show how to find the optimal choice of $\bD$ for a certain relaxation of the problem, after which $\bA$ can be found with a single lattice reduction step. Remarkably, due to the special structure that we stipulate for~$\bA$, the latter problem can be solved much more efficiently than the general case, leading to an algorithm with overall complexity $O(K^3)$, the same as linear precoding methods and lower than previous IF precoding methods \cite{Hong.2012:Reversecomputeforward,Stern.2016:Advancedfactorizationstrategies,He.2018:UplinkDownlinkDualityIntegerForcing}. Simulation results show that the proposed method achieves a higher sum rate than the heuristic choice $\bD=\bI$ and significantly outperforms conventional linear precoding in all simulated scenarios.

\subsubsection*{Notation}

Let $\ZZ$ be the set of integers, and let $\ZZ[j] = \ZZ + \jmath\ZZ$ be the ring of Gaussian integers. The set of all $m \times n$ matrices with entries from the set $\calA$ is denoted as $\calA^{m \times n}$.

\section{Preliminaries} \label{sec:preliminaries}

\subsection{System Model}

Consider a downlink MIMO channel with an $N$-antenna transmitter and $K \leq N$ single-antenna users. Let $\bw_i \in \calW_i$ be the message destined to the $i$th user and $\bx_i \in \CC^n$ be the encoded and modulated version of the message such that $\Ev[\norm{\bx_i}^2] \leq n\SNR$, $i=1,\dotsc,K$, where $\SNR > 0$ is the signal-to-noise ratio and $n$ is the code length. In the following, vectors are treated as row vectors unless otherwise mentioned. Let $\bX = \mat{\bx_1^\tr & \cdots & \bx_K^\tr}^\tr \in \CC^{K\times n}$. After the encoding/modulation, matrix $\bX$ is pre-multiplied by a \emph{precoding matrix} $\bT \in \CC^{N\times K}$, generating the transmitted signals $\bX' = \bT\bX$, where the $j$th row of $\bX'$ is the signal sent be the $j$th transmit antenna, $j=1,\dotsc,N$. The transmitted signals must satisfy an average total power constraint, namely $\Ev[\trace(\bX'\bX'^\herm)] \leq n\SNR$, which always holds if the precoding matrix satisfies $\trace(\bT^\herm\bT) \leq 1.$ 

Let $\by_i \in \CC^{n}$ be the signal received by the $i$th user, $i=1,\dotsc,K$,  and let $\bY = \mat{\by_1^\tr & \cdots & \by_K^\tr}^\tr \in \CC^{K\times n}$. Then, we can express as 
\begin{equation}
\bY = \bH\bX' + \bZ
\end{equation}
where $\bH = \mat{\bh_1^\tr & \cdots & \bh_K^\tr}^\tr \in \CC^{K\times N}$, $\bh_i \in \CC^{n}$ is the channel coefficients to the $i$th user and $\bZ = \mat{\bz_1^\tr & \cdots & \bz_K^\tr}^\tr \in \CC^{K\times n}$ is Gaussian noise, such that $\bz_i \sim \calC\calN(\bfzero,\bI)$.

The $i$th user will try to infer a message $\hat{\bfw}_i \in \calW_i$ from $\bfy_i$. An error occurs if $\hat{\bfw}_i \neq \bfw_i$ for any $i$. The sum rate is given by $\Rsum = R_1 + \cdots + R_K$, where $R_i = \frac{1}{n}\log_2\abs{\calW_i}$. A sum rate $R$ 
is said to be achievable if, for any $\epsilon > 0$ and a sufficiently large $n$, there exists a coding scheme with sum rate at least $R$ and error probability less than~$\epsilon$.

\subsection{Integer-Forcing (IF) Precoding}
\label{sec:IF-precoding}
 
Let $\bA \in \ZZ[j]^{K \times K}$ be a full rank integer matrix. 
For $n$ sufficiently large, there is an IF precoding scheme with achievable sum rate \cite{He.2018:UplinkDownlinkDualityIntegerForcing,Silva.2017:integerforcingprecodingGaussian}
\begin{equation}
R_\IntF(\bfH,\bfA,\bfT) \triangleq \sum_{i=1}^{K} \Rcomp(\bfh'_i,\bfa_i) \label{eq:IF-sum-rate}
\end{equation}
where $\bh'_i \triangleq \bh_i\bT$, $\ba_i$ is the $i$th row of $\bA$,
\begin{equation}
\Rcomp(\bfh_i',\bfa_i) = \log_2^+\left(\frac{1}{\|\bfa_i\|^2 - \frac{1}{\|\bfh_i'\|^2 + \SNR^\inv}|\bfa_i \bfh_i'^\herm|^2}\right).
\end{equation}
is the individual rate for each user, and $\log^+(x) = \max(0,\log(x))$.

Optimal IF precoding consist of finding a matrix $\bA \in \ZZ[j]^{K\times K}$ with $\rank(\bA) = K$ and a matrix $\bT \in \CC^{N \times K}$ with $\trace(\bT^\herm\bT) = 1$ that maximizes \eqref{eq:IF-sum-rate}.

\subsubsection{DIF and RDIF Schemes}

The authors of \cite{Silva.2017:integerforcingprecodingGaussian} proposed two simplified versions of IF precoding, making the problem of finding $\bT$ (and $\bA$) more structured and potentially easier to solve.

The first approach proposed in \cite{Silva.2017:integerforcingprecodingGaussian},  called \emph{diagonally-scaled exact integer-forcing} (DIF) precoding, chooses as precoding matrix $\bT = c\bfH^\herm(\bH\bH^\herm)^\inv\bD\bA,$ where $\bD \in \CC^{K\times K}$ is a diagonal matrix with nonzero entries such that $\abs{\det \bD} =1$ and $c > 0$ is chosen to satisfy $\trace(\bT^\herm\bT) = 1$. 

The DIF precoding is optimal in the high $\SNR$ regime \cite{Silva.2017:integerforcingprecodingGaussian}, where it can achieve a sum rate given by
\begin{equation}
R_\DIF^\HI(\bfH,\bfA,\bfD) \triangleq
K\log_2\left(\frac{\SNR}{\trace\left(\bfA^\herm\bfD^\herm\left(\bfH\bfH^\herm\right)^\inv\bfD\bfA\right)}\right). \label{eq:lower-bound-high-SNR-rate}
\end{equation}

The second approach proposed in \cite{Silva.2017:integerforcingprecodingGaussian}, which is called \emph{regularized DIF} (RDIF), attempts to improve the performance of DIF for finite SNR. Specifically, matrix $\bT$ is chosen as
\begin{equation}
\bfT = c\bfH^\herm\bM\bfD\bfA \label{eq:RDIF_T}
\end{equation}
where
	\begin{equation}
	\bM \triangleq \left(\frac{K}{\SNR}\bfI + \bfH\bfH^\herm\right)^\inv. \label{eq:gram-matrix}
	\end{equation}

The DIF (RDIF) scheme is a generalization of ZF (RZF) precoding which is obtained by making $\bA =\bI$ and $c\bD = \diag(\sqrt{\bp})$, where $\bp \in \RR^K$ is the power allocation vector. Moreover, RDIF reduces to DIF when $\SNR \to \infty$.

\subsection{Problem Statement} \label{sec:problem-statement}

In this paper, we are interested in finding matrices $\bA$ and $\bD$ that maximize the sum rate \eqref{eq:IF-sum-rate} for the RDIF scheme, i.e., with $\bT$ chosen as in \eqref{eq:RDIF_T}. In general, this is a hard problem due not only to the integer constraints on $\bfA$ but also to the complicated objective function \eqref{eq:IF-sum-rate}. The latter difficulty is overcome in \cite{Silva.2017:integerforcingprecodingGaussian} by solving a simpler optimization problem, which can be interpreted as the minimization of a regularized version of the denominator in \eqref{eq:lower-bound-high-SNR-rate}, namely,
\begin{align}
	\minimize_{\bA, \bD}&\quad f(\bfA,\bfD) \triangleq 
	\trace(\bfA^\herm\bfD^\herm\bfM\bfD\bfA) \label{eq:optimization-problem} \\
	\text{s.t.}&\quad \abs{\det \bfD} = 1 \nonumber \\
	&\quad \rank(\bfA) = K \nonumber
\end{align}
where $\bA \in \ZZ[j]^{K \times K}$, $\bD \in \CC^{K \times K}$ is diagonal, and $\bM$ is defined in \eqref{eq:gram-matrix}. While generally a suboptimal heuristic, solving the above problem indeed maximizes the sum rate for the special case of asymptotically high SNR (where RDIF reduces to DIF).

\subsubsection{Special Case of Fixed $\bD$} \label{sec:fixed_D}
If $\bD$ is fixed, then finding $\bA$ that minimizes \eqref{eq:optimization-problem} corresponds to the \emph{shortest independent vector problem} (SIVP) \cite{Silva.2017:integerforcingprecodingGaussian}. Let $\bB^\herm\bB = \bD^\herm\bM\bD$ (i.e, $\bB$ is any square root of $\bD^\herm\bM\bD$). As shown by \cite[Section IV-C]{Silva.2017:integerforcingprecodingGaussian}, we wish to find $K$ shortest linearly independent vectors of the lattice with generator matrix $\bB$ (written in column notation). Those vectors will correspond to the columns of $\bA$. The SIVP can be sub-optimally solved using lattice basis reduction algorithms, for example the Lenstra-Lenstra-Lov\'{a}sz (LLL) algorithm \cite{Lenstra.1982:Factoringpolynomialsrational,Gan.2009:ComplexlatticeReduction}, which has a complexity of $\calO(K^4\log K)$.

When $\bD = \bI$, the RDIF scheme becomes equivalent to the LRA precoding proposed in \cite[eq.~(3)]{Stern.2016:Advancedfactorizationstrategies}, except for the fact that LRA precoding assumes symbol-level detection, while IF precoding
employs codeword-level decoding \cite{He.2018:UplinkDownlinkDualityIntegerForcing,Silva.2017:integerforcingprecodingGaussian}.

\section{Proposed Method} \label{sec:contribution}

In this section we propose a method to find an approximate solution ($\bA,\bD$) to problem \eqref{eq:optimization-problem} for any $K$. 
We start by proposing a convenient choice for the structure of $\bA$.

\subsection{Structure of $\bfA$}

Consider the objective function in \eqref{eq:optimization-problem} and note that
\begin{equation}
f(\bA,\bD) = \sum_{i=1}^K  M_{ii}\norm{\bfa_i}^2\abs{d_i}^2 + \sum_{i=1}^K\sum_{j=i+1}^K 2 M_{ji}\bfa_i\bfa_j^\herm d_id_j^*
\label{eq:objective-function-expanded}
\end{equation}
where $\bfa_i$ is the $i$th row of $\bfA$, $d_i$ is the $i$-th element in the main diagonal of $\bfD$ and $M_{ij}$ is the element of row $i$ and column $j$ of $\bfM$.

The first summation in \eqref{eq:objective-function-expanded} contains only nonnegative values. If we focus exclusively on minimizing $\norm{\ba_i}$, $i=1,\dotsc,K$, then it is easy to see that the optimal choice is $\bA = \bI$. However, since the second summation can have positive or negative values, we wish some degree of freedom to be able to minimize or maximize the absolute values of the inner products ($\abs{\ba_i\ba_j^\herm}$). To satisfy these conflicting requirements, we propose that $\bfA$ be \textit{upper unitriangular} (upper triangular with ones along the main diagonal) up to permutation of rows. An advantage of this structure is that the restriction of full rank $\bA$ is automatically satisfied. Note that, for $K>2$, a row permutation of $\bfA$ may change the achievable rate.

We first consider $\bA$ exactly in upper unitriangular form. The generalization to other permutations is discussed in~\ref{sec:Permutations-of-A}.

\subsection{Relaxed Problem}

Even with the proposed structure for $\bfA$, we still have an integer optimization problem, which is generally hard to solve. In order to circumvent this difficulty, we consider in this section a relaxation of the problem where the indeterminate entries of $\bfA$ can be any complex number.

\begin{theorem} \label{teo:relaxed-problem}
	Under the relaxed constraint that $\bA \in \CC^{K \times K}$ and the additional constraint that $\bA$ be upper unitriangular, problem \eqref{eq:optimization-problem} has a solution given by
	\begin{align}
		\tilA &= \bD^\inv\bfU^\inv\bD = \bLambda^{\frac{1}{2}}\bfU^\inv\bLambda^{-\frac{1}{2}} \label{eq:solution-Atil}\\
		\tilD &= (\det \bLambda)^{\frac{1}{2K}}\bLambda^{-\frac{1}{2}} \label{eq:solution-Dtil}
	\end{align}
	where $\bU \in \CC^{K \times K}$ is upper unitriangular and $\bLambda \in \RR^{K \times K}$ is diagonal such that $\bfM = \bU^\herm\bLambda\bU$. The solution for $\bA$ as a function of $\bD$ is unique and the optimal solution for $\bD$ (with the corresponding optimal $\bA$) is unique up to a phase shift for each of the diagonal entries. 
\end{theorem}
\begin{IEEEproof}
	A proof is given in the Appendix.
\end{IEEEproof}

\subsection{Optimization of $\bA$}

We now show how to find an approximate solution $(\bA,\bD)$ to problem \eqref{eq:optimization-problem} satisfying $\bA \in \ZZ[j]^{K \times K}$, starting from a solution $(\tilA,\tilD)$ to the relaxed problem. First take $\bD = \tilD$, and note that
\begin{align}
	f(\bA,\bD) &= (\det \bLambda)^{\frac{1}{K}} \trace(\bA^\herm\bLambda^{-\frac{1}{2}}\bM\bLambda^{-\frac{1}{2}}\bA) \nonumber\\
	&= (\det \bLambda)^{\frac{1}{K}} \trace(\bA^\herm\tilA^{-\herm}\tilA^\inv\bA) \nonumber \\
	&= (\det \bLambda)^{\frac{1}{K}} \sum_{i=1}^K \norm{\bB\bA(i)}^2 \label{eq:optimal-quantization-problem} 
\end{align}
where $\bB \triangleq \tilA^\inv = \bLambda^{\frac{1}{2}}\bU\bLambda^{-\frac{1}{2}}$ and $\bA(i)$ is the $i$th column of~$\bA$. It follows that finding a Gaussian integer matrix $\bA$ that minimizes \eqref{eq:optimal-quantization-problem} is the same problem described in Section~\ref{sec:fixed_D}. 

\subsection{Permutations} \label{sec:Permutations-of-A}
\newcommand{\Abar}{\bar{\bA}}
\newcommand{\Dbar}{\bar{\bD}}

Let $\Abar$ be an upper unitriangular Gaussian integer matrix and suppose we want to solve \eqref{eq:optimization-problem} under the constraint that $\bA = \bP\Abar$ where $\bP$ is a permutation matrix. 

First, note that 
\begin{align*}
	\trace(\bfA^\herm\bfD^\herm\bfM\bfD\bfA)
	&= \trace(\Abar^\herm \bP^\tr \bfD^\herm \bfM \bfD \bP \Abar) \\
	&= \trace(\Abar^\herm \Dbar^\herm \bP^\tr \bfM \bP \Dbar \Abar)
\end{align*}
where $\Dbar = \bP^\tr \bD \bP$. Thus, we can use the solution of Theorem~\ref{teo:relaxed-problem} with $\bM$ replaced by $\bP^\tr\bM\bP$ to obtain $\bD = \bP\tilD\bP^\tr$ and $\bA = \bP\Abar$, where $\Abar$ is the output of LLL algorithm.

\subsection{Summary of the Method} \label{sec:proposed-summary}

The steps described above allow us to find a choice of $\bA$ and $\bD$ (and thus $\bT$) for any given permutation $\bP$ specifying the structure of $\bA$. A summary of the proposed method is given in Algorithm~\ref{alg:RDIF-Summary}.
\begin{algorithm} 
		\caption{Proposed RDIF Design} 
		\label{alg:RDIF-Summary} 
		\begin{algorithmic}[1]
			\REQUIRE $\bH$ and $\SNR$
			\STATE Compute $\bM = \left(K/\SNR\bI+\bH\bH^\herm\right)^\inv$ \label{step:initialization}
			\STATE Generate a permutation matrix $\bP$ 
			\STATE Compute the LDL decomposition $\bP^\tr\bfM\bfP = \bfU^\herm\bLambda\bfU$ \label{step:LDL-decomposition}
			\STATE Compute $\tilD = \left(\det \bLambda\right)^{\frac{1}{2K}}\bLambda^{-\frac{1}{2}}$ \label{step:tilD}
			\STATE Compute $\bB = \bLambda^{\frac{1}{2}}\bU\bLambda^{-\frac{1}{2}}$ \label{step:tilA}
			\STATE Use the LLL algorithm using $\bB$ as input to find $\Abar$ \label{step:lll-algorithm}
			\STATE Set $\bD = \bP\tilD\bP^\tr$ and $\bA = \bP\Abar$
			\STATE Compute $\bfT_0 \triangleq \bfH^\herm\bfM\bfD\bfA$ \label{step:T0}
			\STATE Compute $c = \trace(\bfT_0^\herm\bfT_0)^{-\frac{1}{2}}$ \label{step:c}
			\STATE Compute $\bfT = c\bfT_0$ \label{step:T}
			\RETURN $\bA$ and $\bT$
		\end{algorithmic}
\end{algorithm}

\subsubsection{Complexity Analysis}

The complexity of an IF scheme is hard to precisely estimate. Generally, the lattice reduction algorithm is the bottleneck on the complexity. It is estimated that the LLL algorithm, one of the most used lattice reduction algorithms, requires $\calO(K^4\log K)$. 	However, in our case, since $\bB$ in step~\ref{step:lll-algorithm} of Alg.~\ref{alg:RDIF-Summary} is an upper unitriangular matrix, the LLL algorithm can be computed with $\calO(K^3)$ \cite{Luk.2008:improvedLLLalgorithm}.
	
Other operations, such as, the computation of matrix $\bM$ in step \ref{step:initialization} or the computation of $\bT$ in steps~\ref{step:T0}-\ref{step:T} require $\calO(NK^2)$ operations each (recall that we assume $N \geq K$). The LDL decomposition in step~\ref{step:LDL-decomposition} requires $\calO(K^3)$ operations. The remaining operations involves only (upper) triangular and diagonal matrices. Therefore, the total complexity is $\calO(NK^2)$, which is the same asymptotic complexity of conventional linear precoding methods.

\section{Simulation Results} \label{sec:results}

In this section we show the average sum-rate performance of the proposed method. In our simulations, the sum rates were obtained through $10000$ channel realizations. In each realization, the channel coefficients were randomly obtained considering a circularly symmetric complex Gaussian distribution with zero mean and unit variance. 

In each simulation, we compare our proposed RDIF design to sum capacity \cite{Tse.2005:FundamentalsWirelessCommunication} and to the conventional linear precoding methods, namely, ZF and RZF. We also compare to the RDIF approach mentioned in Section~\ref{sec:fixed_D}, where we fix $\bD= \bI$ and apply the LLL algorithm to find $\bA$. This method is denoted by ``$\bD = \bI$''.

For our proposed method, we compare two heuristics. Specifically, 
we compare the heuristic where a random permutation is chosen, which is denoted by ``Random'', with a heuristic inspired by \cite{Xu.2012:ParallelCholeskybasedreduction}, where the permutation sorts the diagonal elements of $\bM$ in descending order, which is denoted by ``\mbox{$\bM\!\downarrow$}''. 

\begin{figure}
	\centering
	\includegraphics[width=0.5\textwidth]{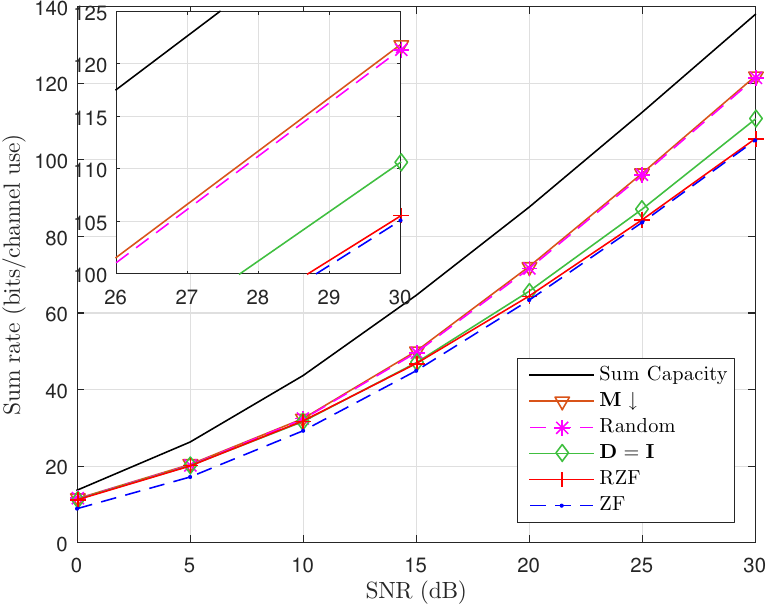}
	\caption{Sum rate for $N=16$ transmit antennas. For each method and each value of $\SNR$, the number of users $K \leq N$ was chosen to maximize the sum rate.  On the box, a close up on $\SNR$ range of $26$ to $30$~dB.}
	\label{fig:SumRateN16}
\end{figure}

Fig.~\ref{fig:SumRateN16}
shows the performance for $N=16$ transmit antennas. For each method and for each value of $\SNR$, we choose, through exhaustive search, the value of $K\leq N$ that achieves the highest sum rate. As expected, the proposed method outperforms linear techniques as well as the previous RDIF approach ($\bD=\bI$) for all values of $\SNR$. In particular, for a sum rate of $105$~bits/channel use, it outperforms the latter by about $2.1$~dB and the former by about $3.2$~dB.

\begin{figure}
	\centering
	\includegraphics[width=0.5\textwidth]{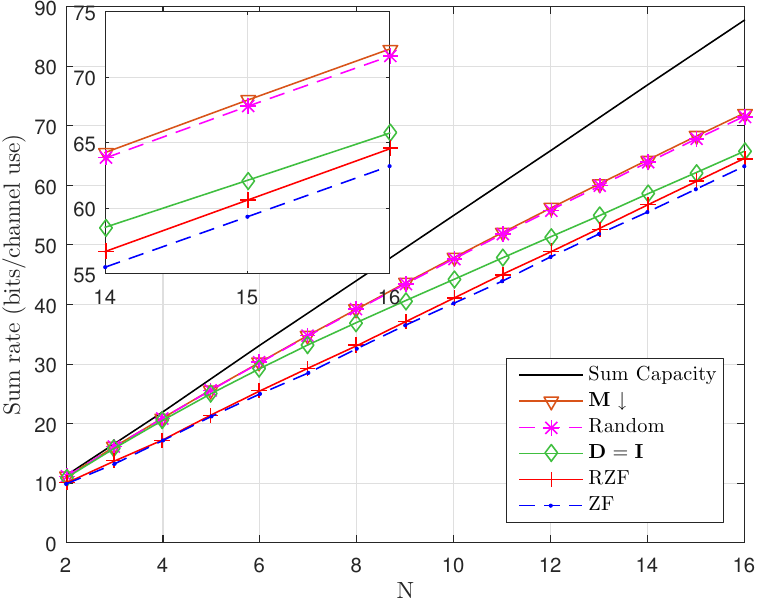}
	\caption{Sum rate for $\SNR = 20$~dB. For each method and each value of $N$, $K$ was chosen to maximize the sum rate. On the box, a close up on the range of $N$ from 14 to 16.}
	\label{fig:Nvarying}
\end{figure}

Fig.~\ref{fig:Nvarying}
shows the performance for a fixed $\SNR=20$~dB while varying the number of transmit antennas $N$ (and again choosing the optimal $K$ for each $N$). Note that, although the gap to capacity increases with $K$, the difference in performance between our proposed method and the other methods considered also increases.

\begin{figure}
		\centering
		\subfloat[]{\includegraphics[width=0.5\textwidth]{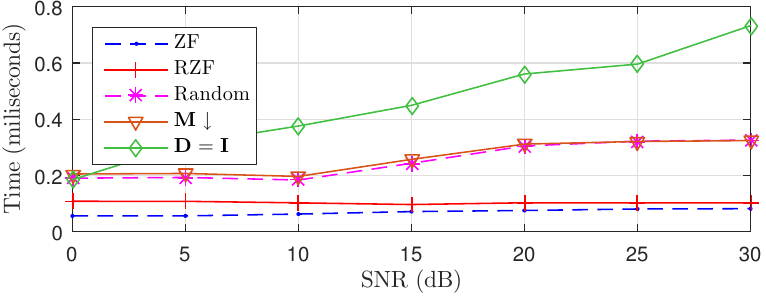}
			\label{fig:N16-simulation-time}}
		
		\subfloat[]{\includegraphics[width=0.5\textwidth]{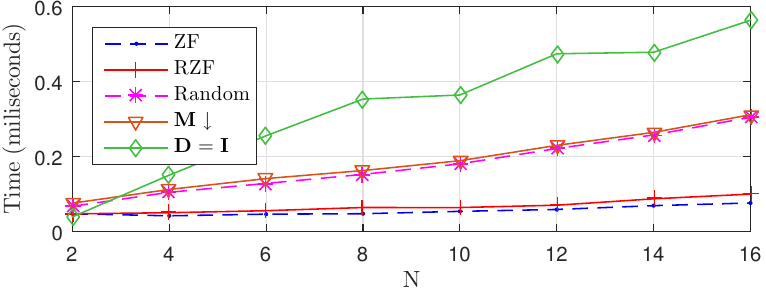}
			\label{fig:SNR20-simulation-time}}
		\caption{Average simulation time for each method. \protect\subref{fig:N16-simulation-time} Parameters as in Fig.~\ref{fig:SumRateN16}.  \protect\subref{fig:SNR20-simulation-time} Parameters as in Fig.~\ref{fig:Nvarying}.}
		\label{fig:time-simulations}
\end{figure}

Fig~\ref{fig:time-simulations}
shows the average time for the simulations of Fig.~\ref{fig:SumRateN16} and Fig.~\ref{fig:Nvarying}. In both situations, we can see that the proposed method is $2$ to $3$ times slower than conventional linear methods. We can also see that the average time of IF methods (the proposed one and $\bD=\bI$) increases with $\SNR$ (and $N$) due to the LLL algorithm. However, since the LLL algorithm is less complex for our proposed method, its simulation time is much smaller than that of $\bD=\bI$ in these scenarios.

Finally, it is worth mentioning that the proposed method for RDIF optimization is indeed suboptimal. As can be seen in Table~\ref{tb:ComparisonK4}, for $N=K=4$, a small but non-negligible gap exists between the performance of our method and that of the exhaustive search carried out in \cite{Silva.2017:integerforcingprecodingGaussian} (which has exponential complexity). Whether this gap can be closed under low complexity is a challenging problem for future work.

\begin{table}
	\caption{Sum rate for $N=K=4$ in bits/channel use.}
	\label{tb:ComparisonK4}
	\centering
	\begin{tabular}{r|c|c|c|c}
		& \multicolumn{4}{|c}{$\SNR$~(dB)} \\
		Method & $0$ & $10$ & $20$ & $30$ \\
		\hline
		Sum capacity & $3.585$ & $10.992$ & $22.071$ & $34.796$ \\
		Exhaustive search \cite{Silva.2017:integerforcingprecodingGaussian} & $3.108$ & $9.970$ & $21.556$ & $34.380$ \\
		Proposed method ($\bM\!\downarrow$) & $3.083$ & $9.884$ & $20.880$ & $33.566$ \\
		\hline 
		Gap from $\bM\!\downarrow$ to \cite{Silva.2017:integerforcingprecodingGaussian} & $0.025$ & $0.086$  &  $0.676$ &   $0.814$	
	\end{tabular}
\end{table}

\section{Conclusion} \label{sec:conclusion}

This letter proposes a low-complexity suboptimal method for RDIF precoding design for $K>2$. The method involves solving a relaxed optimization problem followed by lattice basis reduction in unitriangular form, leading to an overall complexity of $\calO(NK^2)$. Simulation results show that our approach not only significantly outperforms conventional linear precoding, but also improves on previous low-complexity IF precoding both in performance and complexity.

\appendix
\section*{Proof of Theorem~\ref{teo:relaxed-problem}} \label{ap:proof}

Let $\tilA$ and $\tilD$ be a solution to \eqref{eq:optimization-problem} with $\bA \in \CC^{K\times K}$ . We first find $\tilA$ as a function of $\bfD$ and then find $\tilD$. 

Let $\bNabla_{\bfA} f$ be a matrix whose $(i,j)$th element is the partial derivative of \eqref{eq:optimization-problem} with respect to $a_{ij}$ if $i < j$ and zero otherwise. Note that $(\bNabla_{\bfA} f)_{ij} = (2\bfD^\herm\bfM\bfD\bfA)_{ij}$ if $i < j$. The critical points of $f$ with respect to $\bA$ are those which satisfy, for all $j$ and all $i < j$,
\begin{equation}
0 = (\bNabla_{\bfA} f)_{ij} = (2\bfD^\herm\bfM\bfD\bfA)_{ij}.
\end{equation}
Multiplying by $(2d_i^*)^\inv$ and $d_j^\inv$ on both sides, this is equivalent to requiring that, for all $j$ and all $i<j$,
\begin{align}
	0 
	= (\bfM\bfD\bfA\bfD^\inv)_{ij} 
	= (\bfM\bfA')_{ij}  \label{eq:nabla-simplified}
\end{align}
where $\bfA' = \bfD\bfA\bfD^\inv \in \CC^{K\times K}$ is also upper unitriangular. 

Note that \eqref{eq:nabla-simplified} implies that a critical point is any matrix $\bA = \bD^{-1} \bfA' \bD$ such that $\bfM\bfA' = \bL$ is a lower triangular matrix. Thus, any solution, if it exists, can be found by computing an LU decomposition of $\bM = \bL \bA'^{-1}$. Moreover, since we require that the diagonal of $\bA'$ consists of ones, such a decomposition is unique whenever it exists.

Since $\bfM$ is a symmetric positive definite matrix, such an LU decomposition always exists. Specifically, it admits an LDL decomposition $\bfM = \bfU^\herm\bLambda\bfU$, where $\bU$ is an upper unitriangular matrix and $\bLambda$ is a diagonal matrix with real and positive diagonal entries. Thus, $\bfA' = \bU^\inv$ is the unique solution to \eqref{eq:nabla-simplified}, which gives

\begin{equation}
\tilA = \bfD^\inv\bfU^\inv\bfD. \label{eq:optimal-A-proof}
\end{equation}

Now, substituting $\tilA$ in \eqref{eq:optimization-problem}, we have that
\begin{equation}
f(\tilA,\bfD) = \trace(\bfD^\herm\bLambda\bfD) = \sum_{i=1}^K \abs{d_i}^2\lambda_i \label{eq:function_d_relaxed}
\end{equation}
where $\lambda_i > 0$ and $d_i$ are the $i$th diagonal element of $\bLambda$ and~$\bfD$, respectively. Due to the inequality of arithmetic and geometric means, we have that
\begin{equation}
\label{eq:AM-GM-inequality}
\frac{1}{K}f(\tilA,\bfD) = \frac{1}{K}\sum_{i=1}^K \abs{d_i}^2\lambda_i \geq \left(\prod_{i=1}^{K}\abs{d_i}^2\lambda_i\right)^{\frac{1}{K}}
\end{equation}
with equality if and only if $\abs{d_1}^2\lambda_1 = \cdots = \abs{d_K}^2\lambda_K$. 

Applying the constraint $|\det \bD| = 1$, we have
\begin{equation}
\label{eq:AM-GM-inequality-solution}
\left(\prod_{i=1}^{K}\abs{d_i}^2\lambda_i\right)^{\frac{1}{K}} = \left(\prod_{i=1}^{K}\lambda_i\right)^{\frac{1}{K}} = \left(\det\bLambda\right)^{\frac{1}{K}}.
\end{equation}
Thus, the bound in \eqref{eq:AM-GM-inequality} is achievable by setting each term $\abs{d_i}^2\lambda_i$ equal to the right hand side of \eqref{eq:AM-GM-inequality-solution}, i.e.,
\begin{equation}
\bD^\herm\bD = \left(\det \bLambda\right)^{\frac{1}{K}}\bLambda^\inv.
\end{equation}
By choosing each $d_i$ to be real and positive, one solution is given by \eqref{eq:solution-Dtil}, which applied in \eqref{eq:optimal-A-proof} gives \eqref{eq:solution-Atil}.



\begin{thebibliography}{10}
	\providecommand{\url}[1]{#1}
	\csname url@samestyle\endcsname
	\providecommand{\newblock}{\relax}
	\providecommand{\bibinfo}[2]{#2}
	\providecommand{\BIBentrySTDinterwordspacing}{\spaceskip=0pt\relax}
	\providecommand{\BIBentryALTinterwordstretchfactor}{4}
	\providecommand{\BIBentryALTinterwordspacing}{\spaceskip=\fontdimen2\font plus
		\BIBentryALTinterwordstretchfactor\fontdimen3\font minus
		\fontdimen4\font\relax}
	\providecommand{\BIBforeignlanguage}[2]{{%
			\expandafter\ifx\csname l@#1\endcsname\relax
			\typeout{** WARNING: IEEEtran.bst: No hyphenation pattern has been}%
			\typeout{** loaded for the language `#1'. Using the pattern for}%
			\typeout{** the default language instead.}%
			\else
			\language=\csname l@#1\endcsname
			\fi
			#2}}
	\providecommand{\BIBdecl}{\relax}
	\BIBdecl
	
	\bibitem{Tse.2005:FundamentalsWirelessCommunication}
	D.~Tse and P.~Viswanath, \emph{\BIBforeignlanguage{English}{Fundamentals of
			{{Wireless Communication}}}}, 1st~ed.\hskip 1em plus 0.5em minus 0.4em\relax
	{Cambridge, UK ; New York}: {Cambridge University Press}, Jul. 2005.
	
	\bibitem{Bjornson.2014:Optimalmultiusertransmit}
	E.~Bj{\"o}rnson, M.~Bengtsson, and B.~Ottersten, ``Optimal multiuser transmit
	beamforming: {{A}} difficult problem with a simple solution structure,''
	\emph{IEEE Signal Process. Mag.}, vol.~31, no.~4, pp. 142--148, Jul. 2014.
	
	\bibitem{Peel.2005:vectorperturbationtechniquenearcapacitya}
	C.~Peel, B.~Hochwald, and A.~Swindlehurst, ``A vector-perturbation technique
	for near-capacity multiantenna multiuser communication-part {{I}}: Channel
	inversion and regularization,'' \emph{IEEE Trans. Commun.}, vol.~53, no.~1,
	pp. 195--202, Jan. 2005.
	
	\bibitem{Hochwald.2005:vectorperturbationtechniquenearcapacitya}
	B.~Hochwald, C.~Peel, and A.~Swindlehurst, ``A vector-perturbation technique
	for near-capacity multiantenna multiuser communication-part {{II}}:
	Perturbation,'' \emph{IEEE Trans. Commun.}, vol.~53, no.~3, pp. 537--544,
	Mar. 2005.
	
	\bibitem{Avner.2015:vectorperturbationprecoding}
	Y.~Avner, B.~M. Zaidel, and S.~S. Shitz, ``On vector perturbation precoding for
	the {{MIMO Gaussian}} broadcast channel,'' \emph{IEEE Trans. Inf. Theory},
	vol.~61, no.~11, pp. 5999--6027, Nov. 2015.
	
	\bibitem{Windpassinger.2004:Latticereductionaidedbroadcastprecodinga}
	C.~Windpassinger, R.~F.~H. Fischer, and J.~B. Huber, ``Lattice-reduction-aided
	broadcast precoding,'' \emph{IEEE Trans. Commun.}, vol.~52, no.~12, pp.
	2057--2060, Dec. 2004.
	
	\bibitem{Stern.2016:Advancedfactorizationstrategies}
	S.~Stern and R.~F.~H. Fischer, ``Advanced factorization strategies for
	lattice-reduction-aided preequalization,'' in \emph{2016 {{IEEE International
				Symposium}} on {{Information Theory}} ({{ISIT}})}, Jul. 2016, pp. 1471--1475.
	
	\bibitem{Hong.2012:Reversecomputeforward}
	S.-N. Hong and G.~Caire, ``Reverse compute and forward: {{A}} low-complexity
	architecture for downlink distributed antenna systems,'' in \emph{2012 {{IEEE
				International Symposium}} on {{Information Theory Proceedings}}}, Jul. 2012,
	pp. 1147--1151.
	
	\bibitem{He.2018:UplinkDownlinkDualityIntegerForcing}
	W.~He, B.~Nazer, and S.~Shamai~(Shitz), ``Uplink-{{Downlink Duality}} for
	{{Integer}}-{{Forcing}},'' \emph{IEEE Trans. Inf. Theory}, vol.~64, no.~3,
	pp. 1992--2011, Mar. 2018.
	
	\bibitem{Silva.2017:integerforcingprecodingGaussian}
	D.~Silva, G.~Pivaro, G.~Fraidenraich, and B.~Aazhang, ``On integer-forcing
	precoding for the {{Gaussian MIMO}} broadcast channel,'' \emph{IEEE Trans.
		Wirel. Commun.}, vol.~16, no.~7, pp. 4476--4488, Jul. 2017.
	
	\bibitem{Lenstra.1982:Factoringpolynomialsrational}
	A.~K. Lenstra, H.~W. Lenstra, and L.~Lovasz, ``Factoring polynomials with
	rational coefficients,'' \emph{Math. Ann.}, vol. 261, no.~4, pp. 515--534,
	Dec. 1982.
	
	\bibitem{Gan.2009:ComplexlatticeReduction}
	Y.~H. Gan, C.~Ling, and W.~H. Mow, ``Complex lattice {{Reduction Algorithm}}
	for {{Low}}-{{Complexity Full}}-{{Diversity MIMO}} detection,'' \emph{IEEE
		Trans. Signal Process.}, vol.~57, no.~7, pp. 2701--2710, Jul. 2009.
	
	\bibitem{Luk.2008:improvedLLLalgorithm}
	F.~T. Luk and D.~M. Tracy, ``\BIBforeignlanguage{en}{An improved {{LLL}}
		algorithm},'' \emph{\BIBforeignlanguage{en}{Linear Algebra and its
			Applications}}, vol. 428, no. 2-3, pp. 441--452, Jan. 2008.
	
	\bibitem{Xu.2012:ParallelCholeskybasedreduction}
	P.~Xu, ``\BIBforeignlanguage{en}{Parallel {{Cholesky}}-based reduction for the
		weighted integer least squares problem},'' \emph{\BIBforeignlanguage{en}{J
			Geod}}, vol.~86, no.~1, pp. 35--52, Jan. 2012.
	
\end{thebibliography}
\end{document}